\pdfoutput=1

\documentclass[11pt]{article}

\usepackage[]{acl}

\usepackage{times}
\usepackage{latexsym}

\usepackage[T1]{fontenc}

\usepackage[utf8]{inputenc}

\usepackage{microtype}

\usepackage{graphicx}

\usepackage{listings} 

\usepackage{listings} 

\title{DFIN-SQL: Integrating Focused Schema with DIN-SQL for Superior Accuracy in Large-Scale Databases}

\vspace{5cm} 

\author{
  \textbf{Shai Volvovsky} \\
  Sygnia\\
  {\footnotesize\texttt{svolvo@gmail.com}} \\\And
  \textbf{Marco Marcassa} \\
  Safra Bank  \\  
  {\footnotesize\texttt{marcovdkmarcassa@gmail.com}} \\\AND
  \textbf{Mustafa Panbiharwala} \\
  Wells Fargo \\  
  {\footnotesize\texttt{mustafa.shabbir10@gmail.com}}
}

\begin{document}
\maketitle
\begin{abstract}
The task of converting natural language queries into SQL queries is intricate, necessitating a blend of precise techniques for an accurate translation. The DIN-SQL (Decomposed-In-Context SQL) methodology represents a significant development in this domain. This paper introduces DFIN (Decomposed Focused-In-Context), an innovative extension of DIN-SQL that enhances Text-to-SQL conversion by addressing schema linking errors, which are a major source of inaccuracies. DFIN uniquely alternates between prompting techniques and Retrieval-Augmented Generation (RAG), adapting to the size and complexity of the database schema. A preprocessing phase embeds database definitions and leverages annotated files, akin to those in the BIRD dataset, facilitating the runtime retrieval of pertinent schema information. This strategy significantly reduces the token count for schema linking prompts, enabling the use of a standard GPT-4 model over its larger context variant, thus handling large-scale databases more effectively and economically. Our evaluation on the BIRD dataset, a challenging real-world benchmark, demonstrates that DFIN not only scales efficiently but also improves accuracy, achieving a score of 51.69. This improvement surpasses DIN-SQL method (the current third-place), which is the highest-ranked model employing in-context learning rather than fine-tuning, previously scoring 50.72. The advancement of DFIN underscores the evolving capabilities of in-context learning methodologies combined with advanced language models, offering a promising avenue for future research in complex Text-to-SQL conversion tasks.
\end{abstract}

\section{Introduction}

Text-to-SQL conversion stands as a pivotal challenge in the realm of natural language processing. The current state-of-the-art, DIN-SQL (Decomposed-In-Context SQL) \cite{DINSQL}, employs a chained prompt methodology to process natural language queries into SQL commands. Despite its innovative approach, DIN-SQL's \cite{DINSQL} schema linking prompt, which is crucial for identifying relevant schema elements, is prone to errors. These inaccuracies in schema linking account for a significant portion of the total errors encountered. Moreover, as database complexity increases, the context window required by the model expands, leading to higher computational costs and a decrease in accuracy.

\begin{figure}[h]
\centering
\includegraphics[width=\linewidth]{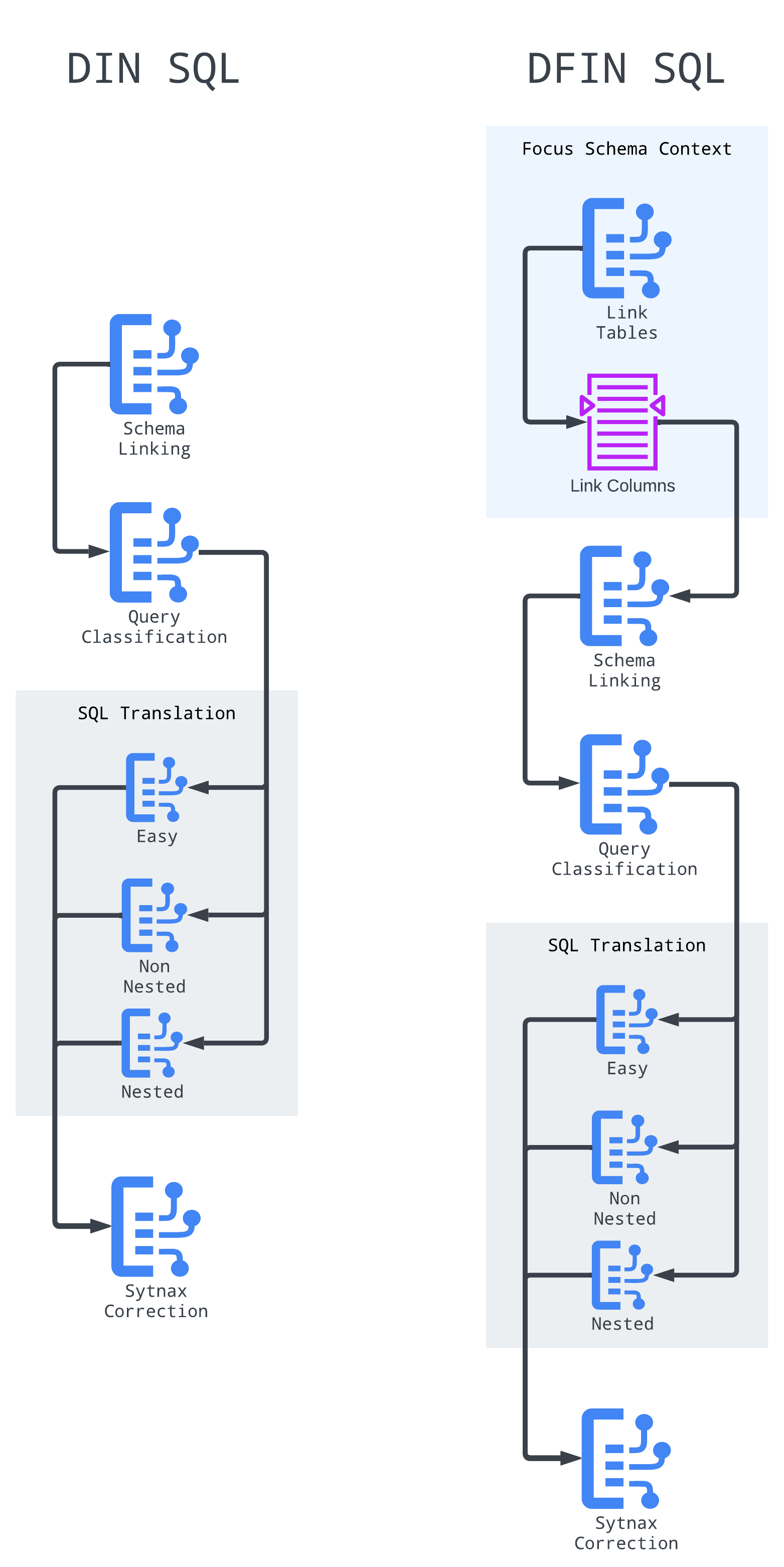} 
\caption{DIN-SQL vs DFIN-SQL.}
\label{fig:dfin}
\end{figure}

In this paper, we propose DFIN-SQL (Decomposed Focused-In-Context), an enhancement to the DIN-SQL \cite{DINSQL} method designed to address these critical issues. Our approach introduces a preliminary focus phase that utilizes a dual strategy, alternating between prompt-based and Retrieval-Augmented Generation (RAG) techniques, to zero in on the most pertinent tables and columns (see Figure \ref{fig:dfin}). This process is dynamic, adjusting according to the complexity and size of the database schema. In our empirical work, prompts were employed to determine table linkages, while the RAG method was applied to refine column selection. This adaptive strategy ensures that DFIN-SQL is capable of handling databases of various sizes, thus optimizing for both precision and recall as necessary.

A key advancement presented in this work is the development of the Schema Linking Accuracy Metric (SLAM), a bespoke metric crafted to assess the precision of the schema focusing phase within DFIN-SQL. By thoroughly preprocessing the development set of the BIRD dataset and aligning our findings with the gold-standard table and column usage stipulated in the queries, we were able to fine-tune our model through iterative refinement. Guided by the SLAM metric, we aimed to achieve a high recall, which is crucial for the next stages in the SQL generation process, while also avoiding the inclusion of superfluous tables and columns that could otherwise lead to inaccuracies in the conversion.

The introduction of DFIN-SQL marks a significant advancement in the accuracy of Text-to-SQL conversion, especially for large and complex database environments. This methodological progress is not merely an incremental improvement but a substantial leap forward, addressing both the scalability and precision challenges faced by current systems. The innovative utilization of preprocessed natural language annotations from the BIRD dataset, alongside the newly established SLAM metric, sets a new standard for precision in Text-to-SQL tasks and signals a promising direction for future exploration in this domain.

\section{Prior Literature}

The research landscape of Text-to-SQL conversion within Natural Language Processing (NLP) is rich and multifaceted. To contextualize our contributions, we categorize relevant literature into two primary groups, each underscoring different aspects of the Text-to-SQL challenge and its solutions.

\subsection{Advances in Text-to-SQL Conversion Mechanisms}

This group encapsulates breakthroughs in conversion methodologies, including model architectures and the prompting mechanisms that guide natural language understanding towards accurate SQL query generation:

\begin{itemize}
  \item \textbf{In-context learning for Text-to-SQL:}  In recent advancements within the field of natural language processing, the text-to-SQL paradigm has experienced a transformative shift with the advent of Large Language Models (LLMs).The C3 technique from \cite{c3} showcases the significance of prompt design in improving LLM performance in a zero-shot context. Similarly, DAIL-SQL \cite{DAILSQL} advances the understanding of prompt engineering, with its Code Representation Prompt (CRP) leading to an 86.6\% accuracy on the Spider \cite{Spider} benchmark. The multi-prompt DIN-SQL method \cite{DINSQL} posits a 'deconstructed-in-context SQL' approach by addressing schema linking and query generation, and recognizes schema linking as the primary source of errors. Our work builds on DIN-SQL's \cite{DINSQL} framework, focusing on refining its application to large databases where efficiency and accuracy are key.
  
  \item \textbf{Prompting and De-semanticization Techniques:}  Innovations in Text-to-SQL conversion have led to novel strategies that refine query generation through advanced prompting techniques. The study by \cite{de-semantic} introduces a de-semanticization process paired with skeleton retrieval, stripping natural language questions to their structural core to enhance the accuracy of SQL translations performed by GPT-3.5. This approach underscores the significance of domain-specific knowledge and iterative SQL revisions in the context of varied datasets. Complementing this, \cite{RAG} advances a retrieval-augmented prompting framework that employs Large Language Models for iterative prompting and question simplification. This method dynamically adapts to feedback, iteratively refining SQL queries to approach the desired result more accurately. Both methods, showcasing improved performance on benchmark datasets, suggest that the integration of these structured prompting techniques could significantly bolster the development of more precise and efficient Text-to-SQL models.
\end{itemize}

\subsection{Datasets and Schema Understanding in Text-to-SQL}

The development of specialized datasets and the comprehension of database schemas are critical for advancing Text-to-SQL technologies. Here we delve deeper into the BIRD dataset which plays a central role in our work:

\begin{itemize}
  \item \textbf{BIRD - A BIg Bench for Database Grounded Text-to-SQLs:} The BIRD dataset's \cite{BIRD} intricate annotations of database schemas distinguish it as a pivotal resource for our study. It provides extensive natural language descriptions for columns, essential for the preprocessing step in our DFIN-SQL approach, and offers a more realistic challenge for models due to its complexity and domain-specific data.
  
  \item \textbf{Foundational Benchmarks:} The SPIDER dataset \cite{Spider}, while foundational, presents a less challenging benchmark and lacks the detailed annotations found in BIRD \cite{BIRD}. The BIRD dataset \cite{BIRD}, with its comprehensive annotations and our own Schema Linking Accuracy Metric (SLAM), pushes the envelope in terms of precision for Text-to-SQL tasks and reflects a more accurate representation of real-life database scenarios.
\end{itemize}

Our DFIN-SQL methodology synthesizes these advances, presenting a novel approach that significantly enhances accuracy for complex and large database environments, marking a notable progression in the field of Text-to-SQL conversion.

\section{Data}

\begin{table*}[!htbp]
  \centering
  \caption{BIRD Dev Databases}
  \label{tab:dev_database_description}
  \resizebox{\textwidth}{!}{%
  \begin{tabular}{|c|c|c|c|c|}
    \hline
    \textbf{Database Name} & \textbf{\# of Tables} & \textbf{Median \# of Columns} & \textbf{Range of Columns (Min-Max)} \\ \hline
    California Schools & 3 & 29 & 11-49 \\
    Card Games & 6 & 6 & 4-74 \\
    Codebase Community & 8 & 6 & 4-21 \\
    Debit Card Specializing & 5 & 3 & 2-9 \\
    European Football & 7 & 7 & 2-115 \\
    Financial & 8 & 5 & 4-16 \\
    Formula 1 & 13 & 7 & 2-18 \\
    Student Club & 8 & 6 & 2-9 \\
    Superhero & 10 & 2 & 2-12 \\
    Thrombosis Prediction & 3 & 13 & 7-44 \\
    Toxicology & 4 & 3 & 2-3 \\ \hline
  \end{tabular}%
  }
\end{table*}

Our evaluation was conducted on the cross-domain BIRD dataset \cite{BIRD} which consists of a train and dev set. For our analysis and work, we primarily focus on the dev split as the test split is not publicly available. The dev dataset contains 1533 question-SQL pairs over 11 databases. Each instance comprises a natural language question on a specific database and its corresponding SQL query. The descriptions of the 11 databases are presented in Table \ref{tab:dev_database_description}. Each database is meticulously curated, with foreign key constraints and calibrated names for abbreviated columns or tables. Furthermore, a unique "Table Description File" was formulated for each table across all databases to aid annotators in grasping the structure of databases, offering insights into full schema names and the significance of database values. These files are pivotal for models to discern intricate database structures and comprehend values which might not string-match directly with questions. Each database file also contains an individual SQLite file for the database, which includes the schema, data, and metadata for each of the tables.

Additionally, the BIRD dataset \cite{BIRD} provides extensive annotations for each column within the databases. An example annotation from the \texttt{frpm} table in the \texttt{california schools} database is as follows:

\begin{itemize}
  \item \textbf{Original Column Name:} \\
  \texttt{'FRPM Count (K-12)'}
  \item \textbf{Column Description:} \\
  Free or Reduced Price Meal Count (K-12)
  \item \textbf{Data Format:} \\
  real
  \item \textbf{Value Description:} \\
  Commonsense evidence: eligible FRPM rate = FRPM / Enrollment
\end{itemize}

Apart from the databases, the dev dataset also comes with a \texttt{dev.json} file which contains the gold standard 1533 question-SQL pairs. The \texttt{dev.json} file has fields including \texttt{question\_id}, \texttt{db\_id}, \texttt{question}, \texttt{evidence}, \texttt{SQL}, and \texttt{difficulty}. The \texttt{evidence} field represents external knowledge given to both the model and the annotators to improve their ability to answer the question correctly. The \texttt{difficulty} field classifies all SQL queries as simple, moderate, or challenging. The annotators have calculated the difficulty of a task based on factors such as question comprehension, schema linking, external knowledge acquisition, and reasoning. The distribution of queries based on difficulty is depicted in Table \ref{tab:question_sql_pairs}.

\begin{table}[h!]
  \centering
  \begin{tabular}{|c|l|r|}
    \hline
    Difficulty & \# of question-sql pairs & Count \\ \hline
    1 & Simple & 925 \\
    2 & Moderate & 465 \\
    3 & Challenging & 144 \\ \hline
  \end{tabular}
  \caption{Number of Question-SQL Pairs by Difficulty}
  \label{tab:question_sql_pairs}
\end{table}

A sample entry from the \texttt{dev.json} file:

\begin{itemize}
  \item \textbf{question\_id}: 0
  \item \textbf{db\_id}: california\_schools
  \item \textbf{question}: What is the highest eligible free rate for K-12 students in the schools in Alameda County?
  \item \textbf{SQL}:\texttt{SELECT `Free Meal Count (K-12)` / `Enrollment (K-12)` FROM frpm WHERE `County Name` = 'Alameda' ORDER BY (CAST(`Free Meal Count (K-12)` AS REAL) / `Enrollment (K-12)`) DESC LIMIT 1}
  \item \textbf{evidence}: Eligible free rate for K-12 = `Free Meal Count (K-12)` / `Enrollment (K-12)`
  \item \textbf{difficulty}: simple
\end{itemize}

\section{Methodology}
\label{sec:methodology}

The DFIN-SQL methodology arises from the need to enhance schema linking in SQL query generation, where the standard DIN-SQL \cite{DINSQL} approach often incorporates an unwieldy and overgeneralized context. DIN-SQL \cite{DINSQL} contexts include a \texttt{CREATE TABLE} statement, sample table rows, and column descriptions for every table in a database, which creates a bulky and sometimes irrelevant schema context for the task at hand. This broad context can dilute the model's focus, leading to suboptimal schema linking performance. An illustrative example of this standard context is presented in Figure \ref{fig:employee_schema_info}, showcasing the typical structure and content included.

\begin{figure*}[ht]
\centering
\begin{lstlisting}[language=SQL, basicstyle=\ttfamily\footnotesize, frame=single]
CREATE TABLE employee (
    emp_id INT PRIMARY KEY,
    first_name VARCHAR(100),
    last_name VARCHAR(100),
    gender CHAR(1),
    department VARCHAR(100)
);
-- Sample rows from 'employee' table:
-- emp_id | first_name | last_name | gender | department
-- 101    | John       | Doe       | M      | Engineering
-- 102    | Jane       | Smith     | F      | Marketing
-- 103    | Emma       | Brown     | F      | Human Resources

-- Column descriptions:
-- 'emp_id': Unique identifier for each employee.
-- 'first_name': Given name of the employee.
-- 'last_name': Family name of the employee.
-- 'gender': Gender of the employee ('F' for female, 'M' for male).
-- 'department': Department where the employee works.
\end{lstlisting}
\caption{Example of a single table representation for the Schema Linking prompt}
\label{fig:employee_schema_info}
\end{figure*}

\subsection{Preprocessing for Table Descriptions}
To address the absence of overarching table descriptions in the BIRD dataset \cite{BIRD}, we utilized GPT-4 to generate succinct and informative descriptions for each table. This was achieved by prompting the model with a structured input that included the table's \texttt{CREATE TABLE} statement, sample rows, and existing column descriptions, analogous to the process depicted in Figure \ref{fig:employee_schema_info}. The resulting table descriptions were then stored in a 'table\_description.json' file for subsequent use during the schema linking process.

\subsection{Runtime Processing}
\subsubsection{Linking Tables}
At runtime, when a natural language question is posed, DFIN-SQL employs a prompt-based method to determine the relevant tables from the database. Two strategies are employed: a 'minimal' approach, which seeks to identify only the most critical tables, and a 'conservative' approach, which aims for higher recall by including any potentially relevant tables. Examples of these prompts are presented in Figures \ref{fig:minimal_prompt} and \ref{fig:conservative_prompt}.

\begin{figure*}[ht]
\centering
\begin{lstlisting}[language=Python, basicstyle=\ttfamily\footnotesize, frame=single]
"""You are a sql query assistant schema linker. You are provided with a natural language question a possible hint and a list of table descriptions.
Your task is to pick the minimum relevant tables for the query. Each table you pick is costly so be cautious and weary for each table you consider.
the final output of tables should look like a pytho set -> {{'table_a', 'table_b'}}

Table Descriptions:
####

{tables_descriptions}

####

Question: {question}
{f"Hint: {hint}" if hint else ''}

tables for query, Let's think step by step."""
\end{lstlisting}
\caption{Example of a minimal table linking prompt}
\label{fig:minimal_prompt}
\end{figure*}

\begin{figure*}[ht]
\centering
\begin{lstlisting}[language=Python, basicstyle=\ttfamily\footnotesize, frame=single]
"""You are a sql query assistant schema linker with a strong focus on high recall. You are provided with a natural language question a possible hint and a list of table descriptions.
Your task is to identify all the tables that could possibly be relevant to the query. It is crucial that you do not miss any relevant tables, even if it means including a few that might not be strictly necessary. Each missed table is a significant issue, while including an extra table is a minor inconvenience.

Table Descriptions:
...

Considering the importance of high recall and the relative cost of missing a table versus including an extra one, list the tables for the query. Aim for completeness and be err on the side of inclusion. Your output should be formatted as a python set -> {{'table_a', 'table_b'}}.

Tables for query, Let's think step by step.
\end{lstlisting}
\caption{Example of a conservative table linking prompt}
\label{fig:conservative_prompt}
\end{figure*}

\subsubsection{Linking Columns}
In the DFIN-SQL approach, once the pertinent tables are pinpointed, an embedding similarity step hones the context further by pinpointing columns with the greatest semantic relation to the given query. This step is strictly applied to the tables identified previously, ensuring relevance and efficiency.

After preprocessing all column description embeddings, generated using the OpenAI model ada-002, are compared to the question's embedding through cosine similarity. The top \( k \) columns with the highest similarity are selected, focusing the query formation on the most relevant columns. This method effectively narrows down the necessary columns for constructing the SQL query.

Moreover, regardless of their cosine similarity scores, primary and foreign keys from the linked tables are always included to accommodate any join operations required by the query.

\subsection{Final Context Construction}
The final stage in the DFIN-SQL methodology involves assembling a concise context that encompasses only the crucial schema information as identified in the previous steps. This focused context is then utilized to guide the language model in generating an accurate SQL query, reflecting the specific requirements of the natural language question posed.


\section{Experiment}

\subsection{Models}
Our choice of models for the experiments includes GPT-4 \cite{GPT4} for its unparalleled natural language processing capabilities and OpenAI's ADA2 model for generating embeddings used in schema linking. The combination of GPT-4's \cite{GPT4} generative abilities and ADA2's efficient embeddings allows us to capitalize on the strengths of both models to improve the overall accuracy of SQL generation from natural language queries.

\subsection{Evaluation Metrics}
Our evaluation framework employs Execution Accuracy (EX) and Valid Efficiency Score (VES) as principal metrics, following the benchmarks established by the creators of the BIRD dataset \cite{BIRD}. Execution Accuracy (EX) is a critical metric that measures the proportion of SQL queries generated by the model that produce the same results as the ground truth when executed against the database. This metric directly reflects the model's ability to understand and translate a user's natural language question into a functionally equivalent SQL query.

In parallel, the Valid Efficiency Score (VES) quantifies the efficiency of the SQL queries generated. It is computed by considering only those queries that are both syntactically correct and yield the correct result set when executed. The VES offers insight into the computational cost of the generated queries, encouraging the model to produce optimized SQL statements that align with the intended user requests while minimizing resource utilization.

Together, EX and VES provide a dual perspective on the model's performance, ensuring that the generated SQL queries are not only accurate in representing the user's intent but are also efficient in their execution. This holistic approach to evaluation aligns with our goal of improving the functionality of SQL generation from natural language, ensuring that the generated queries are practical for real-world applications where accuracy and efficiency are paramount.

\subsection{Schema Linking Accuracy Metric (SLAM)}

The Schema Linking Accuracy Metric (SLAM) is a pivotal element of our evaluation framework, enabling us to meticulously assess the precision and recall of our schema linking method. SLAM operates by contrasting the predicted linked tables and columns against a gold standard—a preprocessed set of tables and columns that are known to be relevant to each query within the dataset. The analysis is twofold: it first evaluates the table linking accuracy by comparing the predicted set of relevant tables with the gold standard, tracking metrics such as missed and extra tables. It then delves into the granularity of column linking, comparing predicted columns within these tables to the gold standard, accounting for the possibility of a greater number of columns being linked due to the top-k selection approach, which ensures that even tables with fewer than k columns have all columns considered. The metric focuses particularly on recall, especially for columns, to minimize false negatives and ensure comprehensive coverage of relevant schema elements. This is crucial for tasks that involve information retrieval, where missing relevant columns (false negatives) can significantly impact the accuracy and utility of the results. The SLAM's output includes detailed per-question recall rates as well as aggregate statistics, such as average recall across the dataset, which informs further refinement of the linking process and prompt optimization.

\subsection{Implementation}
In the implementation of our methodology, the embeddings generated by ADA2 play a pivotal role in the identification of semantically relevant columns, allowing us to employ cosine similarity measures for precise schema linking. Our focused context approach enabled us to utilize the GPT-4-8k model variant, which is less resource-intensive than the GPT-4-32k variant used in the original DIN-SQL \cite{DINSQL} study, while improving overall accuracy and efficiency.

\subsection{Baselines and Permutations}
Our comparative analysis involved benchmarking against the existing DIN-SQL \cite{DINSQL} framework and the base GPT-4 \cite{GPT4} model, drawing from the BIRD dataset's \cite{BIRD} development set. To fine-tune our schema linking strategy, we initially focused on this stage exclusively, deploying our Schema Linking Accuracy Metric (SLAM) to assess a series of permutations. These included both minimal and conservative modes with top-k settings of 15, 10, and 5 for column selection. Through this process, we were able to rapidly iterate and evaluate the impact of each configuration on schema linking accuracy. Ultimately, we settled on the minimal mode with a top-k of 15 for the full DFIN-SQL integration. This decision was informed by the observation that while the conservative approach did yield improvements in recall, the overall gains did not justify the additional complexity. The chosen configuration, therefore, struck a harmonious balance between precision and recall, leading to a significant enhancement in the accuracy of the final query predictions.

\begin{figure}[ht]
\centering
\includegraphics[width=\linewidth]{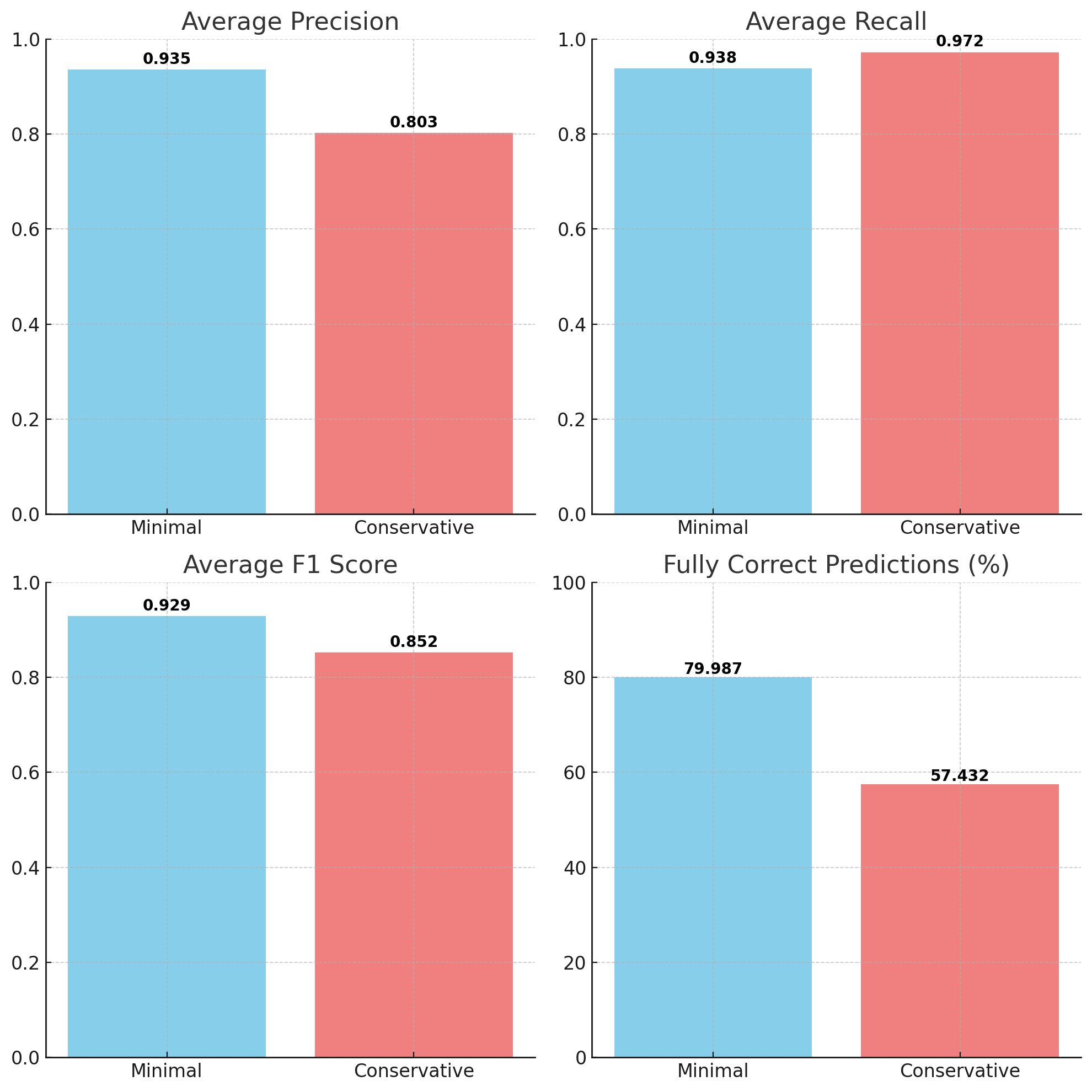}
\caption{Analysis of table links.}
\label{fig:table_links_analysis}
\end{figure}

\begin{figure}[ht]
\centering
\includegraphics[width=\linewidth]{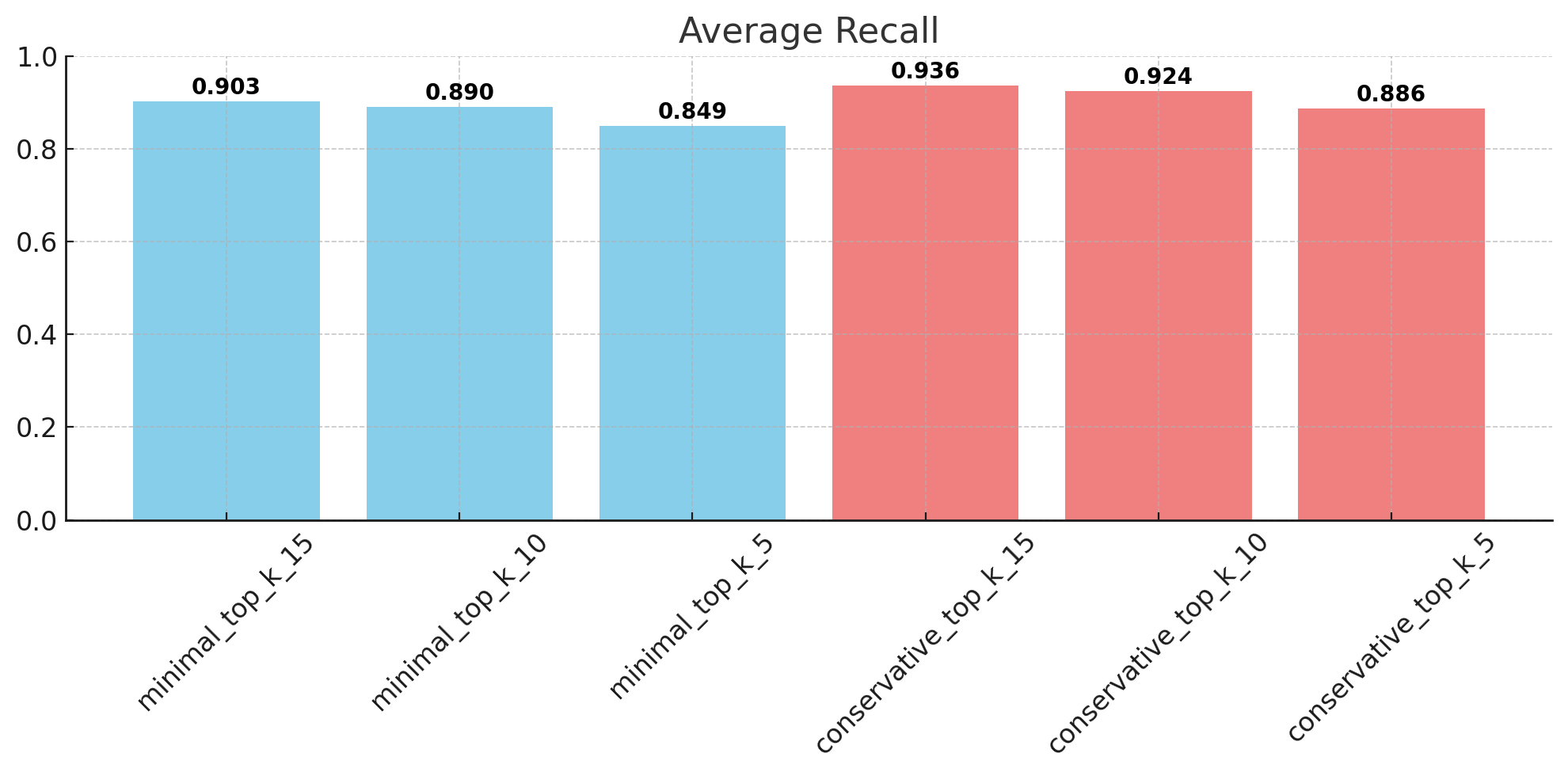}
\caption{Recall of column links.}
\label{fig:column_links_recall}
\end{figure}

\section{Analysis}

\begin{figure*}[tb] 
\centering
\includegraphics[width=\textwidth]{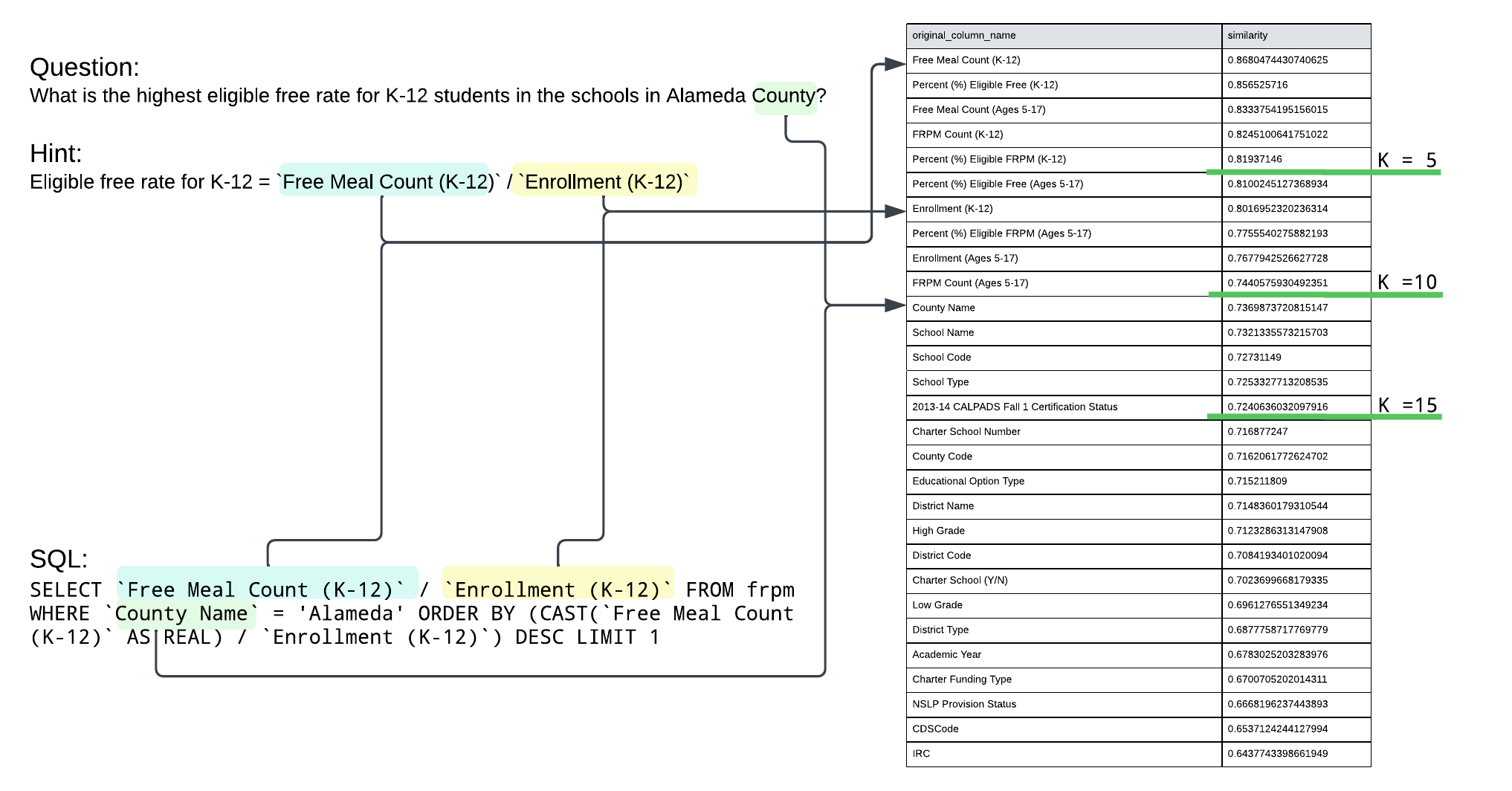}
\caption{Example Top-K Distribution for a Question. The relevance of columns is plotted, showing the uniform distribution and highlighting the selected relevant columns.}
\label{fig:top_k_distribution}
\end{figure*}

\subsection{Schema Focusing Impact on Accuracy and Efficiency}
Our comprehensive analysis indicates a significant enhancement in both accuracy and efficiency when the schema relevant to the query is focused within the schema linking prompt. This refinement of the schema is particularly beneficial when dealing with large databases, a common scenario in real-world applications. Such databases can contain hundreds of tables and columns, posing a substantial challenge for traditional natural language processing methods. By distilling the schema to the most pertinent elements, our methodology demonstrates robustness and scalability, enabling successful handling of complex database structures without compromising on performance.

\subsection{Table Linking Precision and Recall}
The analysis of table links brings to light a key finding: the minimal version of our approach yields an average precision of 0.935, significantly surpassing the conservative approach, which stands at 0.803. Even though the conservative method slightly edges out with an average recall of 0.972 compared to the minimal approach's 0.938, the F1 score—a measure of a test's accuracy—and fully correct predictions favor the minimal approach. Impressively, in 80\% of the questions, the table linking prompt correctly identified the relevant tables while filtering out irrelevant ones. This high percentage of fully correct predictions underscores the effectiveness of our minimal approach, striking an optimal balance between inclusivity and precision.

\subsection{Top-K Column Selection Insights}
An intriguing aspect of our method is the top-K column selection. A visualization, which we will refer to as Figure~\ref{fig:top_k_distribution}, exemplifies the challenge of selecting an appropriate K. In the figure, the distribution of column relevance for a given question shows that the relevant column for the SQL query is ranked at position 11, indicating the risks associated with selecting a small top-K. The distribution of similarities across the columns is relatively uniform, complicating the task of distinguishing between highly relevant and less relevant columns.

This uniform distribution prompts a reconsideration of our exclusive reliance on the ADA-002 embeddings for semantic representation. While embeddings provide a robust baseline for relevance, our findings suggest that integrating a re-ranking mechanism that incorporates exact keyword matches may yield better results. A hybrid approach combining semantic embeddings with keyword matching could allow for a reduction in the top-K threshold while maintaining, or even improving, the precision of column selection. This re-ranking could be particularly advantageous for queries that naturally involve a limited number of columns, aligning the focus context more closely with the actual usage patterns observed in SQL queries.

\subsection{Rethinking the Top-K Threshold}
The analysis suggests that as the number of columns increases, the strategy for selecting the top-K columns warrants a re-evaluation. Our objective is to confine the focus context to the relevant columns, avoiding the inclusion of superfluous information. However, the necessity of including crucial columns for accurately interpreting the query at times necessitates a larger top-K. The future development of our methodology may involve exploring a more nuanced column selection mechanism that can dynamically adjust the top-K based on the distribution of semantic similarities and the presence of keyword matches. This adjustment would optimize the relevance of the included columns, thus enhancing the overall effectiveness of the schema linking process.

\subsection{Insights on Element Size and Method Suitability}

The size of the element set in database queries significantly impacts the effectiveness of different schema focusing techniques. Our analysis reveals that when dealing with a smaller set of elements, such as tables where the average database contains between 8 to 13 tables, a direct prompting strategy yields superior results. This technique involves the integration of table descriptions within the prompts used by GPT-4 \cite{GPT4}, capitalizing on the model's language understanding capabilities to determine relevance.

Conversely, as the number of elements increases, the viability of direct prompting diminishes. In cases where databases contain an extensive number of columns, such as the European football database with its 115 columns, it is impractical to incorporate all column descriptions within a single prompt. Here, our Retrieval-Augmented Generation (RAG) approach proves to be more effective. It enables the model to handle a larger element set by leveraging embeddings to distill a focused context. This approach, when paired with a carefully calibrated top-k threshold and a re-ranking mechanism that accounts for keyword matches, can significantly enhance the model's ability to discern the most relevant columns for a given query.

The dichotomy between the two scenarios emphasizes the importance of method adaptability based on element set size. Direct prompting excels in smaller, more manageable scenarios, while the RAG approach, augmented with strategic re-ranking, is better suited for larger, more complex element sets. This insight lays the groundwork for future research and development, suggesting that a hybrid approach or a dynamic method selection based on the size of the element set could further optimize the schema focusing process.

\section{Conclusion}

In this work, we have presented DFIN-SQL, an enhancement over the DIN-SQL \cite{DINSQL} approach, which addresses the critical issue of schema linking errors in Text-to-SQL conversion tasks. By introducing a novel preprocessing phase and a dynamic runtime processing mechanism, DFIN-SQL significantly reduces the token count for schema linking prompts, thereby allowing the standard GPT-4 \cite{GPT4} model to handle large-scale databases more effectively. Our methodological innovations, including the Schema Linking Accuracy Metric (SLAM) and the use of Retrieval-Augmented Generation (RAG), have demonstrated a marked improvement in the accuracy and scalability of SQL query generation from natural language descriptions.

The empirical results on the BIRD dataset underscore the potential of DFIN-SQL to achieve superior performance on complex, real-world databases. By focusing on the schema elements that are most pertinent to the query, DFIN-SQL not only streamlines the query generation process but also provides a more economical solution in terms of computational resources.

\subsection{Refinement and Future Directions}

Our study on DFIN-SQL demonstrates that a focused schema can significantly elevate the precision of Text-to-SQL translations. Turns out, in the realm of database querying, focus is all you need :) The use of a minimal approach for table linking and a top-k strategy for column selection has proven to be effective, particularly when dealing with the intricate schemas of large databases. However, our analysis indicates that there is potential for further optimization. A hybrid strategy that combines semantic relevance with keyword matching could refine the column selection process, particularly for databases with numerous columns.

The distinction between when to use direct prompting and when to employ RAG-based techniques is a key takeaway from our research. For smaller element sets, direct prompting is practical and efficient, but as the size of the element set grows, RAG becomes necessary to manage the increased complexity. Adjusting the top-k threshold dynamically according to the distribution of semantic similarities could further enhance our methodology.

In conclusion, DFIN-SQL represents a significant step forward in the field of Text-to-SQL conversion. Our findings lay a foundation for future work, which will explore more sophisticated selection mechanisms and the potential of a dynamic approach that tailors the methodology to the size of the database schema. The path ahead is promising, with the ultimate goal of creating models that seamlessly bridge the gap between natural language and SQL for databases of any scale.




\bibliographystyle{acl_natbib}
\bibliography{main}

\end{document}